\documentstyle[11pt,newpasp,twoside,psfig]{article} 
\def\edcomment#1{\iffalse\marginpar{\raggedright\sl#1\/}\else\relax\fi} 
\reversemarginpar 

\begin{document} 
\title{Formation of Globular Clusters in Merging Galaxies}

\author{Fran\c cois Schweizer}
\affil{Carnegie Observatories, 813 Santa Barbara Street, Pasadena, \\
CA 91101-1292, USA}

\begin{abstract} 
Collisions and mergers of gas-rich galaxies trigger bursts of star and
cluster formation.  Of the thousands of clusters typically formed during
a major merger, only the most massive and compact survive for
Gigayears as globular clusters (GCs).  In $\la$1~Gyr old merger remnants,
these `second-generation' GCs appear by the hundreds as young halo clusters
of $\sim$\,solar metallicity.  Their likely descendants, metal-rich GCs of
intermediate age (2\,--\,5~Gyr), have recently been found in $\sim$10 E
galaxies, where they appear as slightly overluminous red GCs with a still
power-law-like luminosity function.  Their color and radial distributions
suggest that they evolve into the red metal-rich GCs observed in old
ellipticals.  There is good evidence that second-generation GCs form from
giant molecular clouds shocked by the rapid pressure increase in
merger-induced starbursts.  This mechanism supports the view that the
universal pressure increase during cosmological reionization may have
triggered the formation of the metal-poor globulars observed in galaxies
of all types.
\end{abstract}

\section{Homage to Ivan King}

Ivan King was my revered teacher.  I owe him a lot, including an interest in
globular clusters. Into my copy of his textbook {\it The Universe Unfolding}
(1976), he wrote the dedication:  ``Make as much of this obsolete as you can,
Fran\c cois! Best wishes, Ivan.''  In this spirit I put forth the hypothesis
that major mergers of gas-rich spirals form not only ellipticals, but
also new globular clusters within them (Schweizer 1987).  I cherish Ivan's
later comment on this hypothesis: ``When Schweizer suggested resolving the
$S_N$ problem by making globular clusters in mergers, I thought
it was ridiculous.  But now in the HST observations of NGC 1275 we seem to
see it actually happening'' (King 1993).  

Always quick to grasp major issues, Ivan challenged Alar Toomre at the
Yale Conference as follows (King 1977):
``You showed us 10 merging pairs [of galaxies] and then asked us to look
for, or at least accept the existence of, 500 remnants from so long ago
that they no longer bear the `made by Toomre' label.  I would be much
more impressed if you showed us the 20 or 30 systems in the box immediately
adjacent in your histogram.  What do these merged pairs look like in their
next few galactic years?''

As I hope to show in the present review, the
descendants of major disk--disk mergers---i.e., ellipticals with globular
clusters of intermediate age---are now being found in growing numbers.

\section{Cluster Formation in Ongoing Mergers}

Collisions and mergers of gas-rich spirals trigger bursts of intense star
and cluster formation. Some well-known examples of ongoing mergers with
spectacular systems of young clusters are NGC 4038/39 (Whitmore \& Schweizer
1995; Whitmore et al.\ 1999), NGC 3256 (Zepf et al.\ 1999), and NGC 6052
(= Mrk 297, Holtzman et al.\ 1996).  Although such mergers form star
clusters by the thousands, it is currently difficult to predict what
fraction of the clusters will survive as globular clusters (hereafter GC).
Presumably only the more massive and compact clusters will survive for
several Gyr, while many more fragile clusters and associations will disperse
within a few internal crossing times.  Nevertheless, several important
results have emerged from {\it Hubble Space Telescope (HST)} studies of
the above systems, and especially
of NGC 4038/39 (Fig.~\ref{fig1}):  (1) Star clusters tend to form in regions
of high gas density and are, therefore, clustered themselves; typically
10--20 young clusters belong to a complex previously identified from the
ground as a giant H$\;$II region.  (2) To a good approximation the cluster
luminosity function is a power law,  $\phi(L)dL \propto L^{-\alpha} dL$,
with $1.7 \la \alpha \la 2.1$ and no evidence of any turnover at fainter
magnitudes.  (3) The most luminous clusters clearly show properties to be
expected of massive young GCs and appear to display signs of structural
evolution as a function of their age.

\setcounter{figure}{0}
\begin{figure}[t]
\centerline{\hbox{\vsize 6.5truecm
\psfig{figure=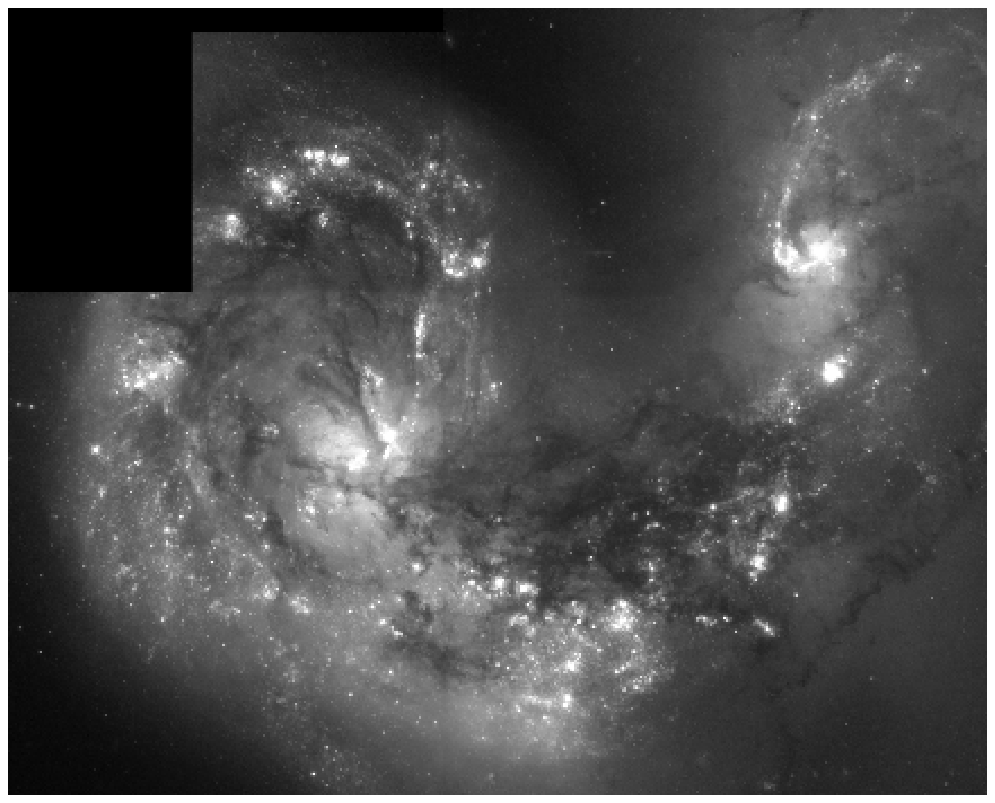,height=6.5cm,angle=270}
\hskip 0.5truecm
\psfig{figure=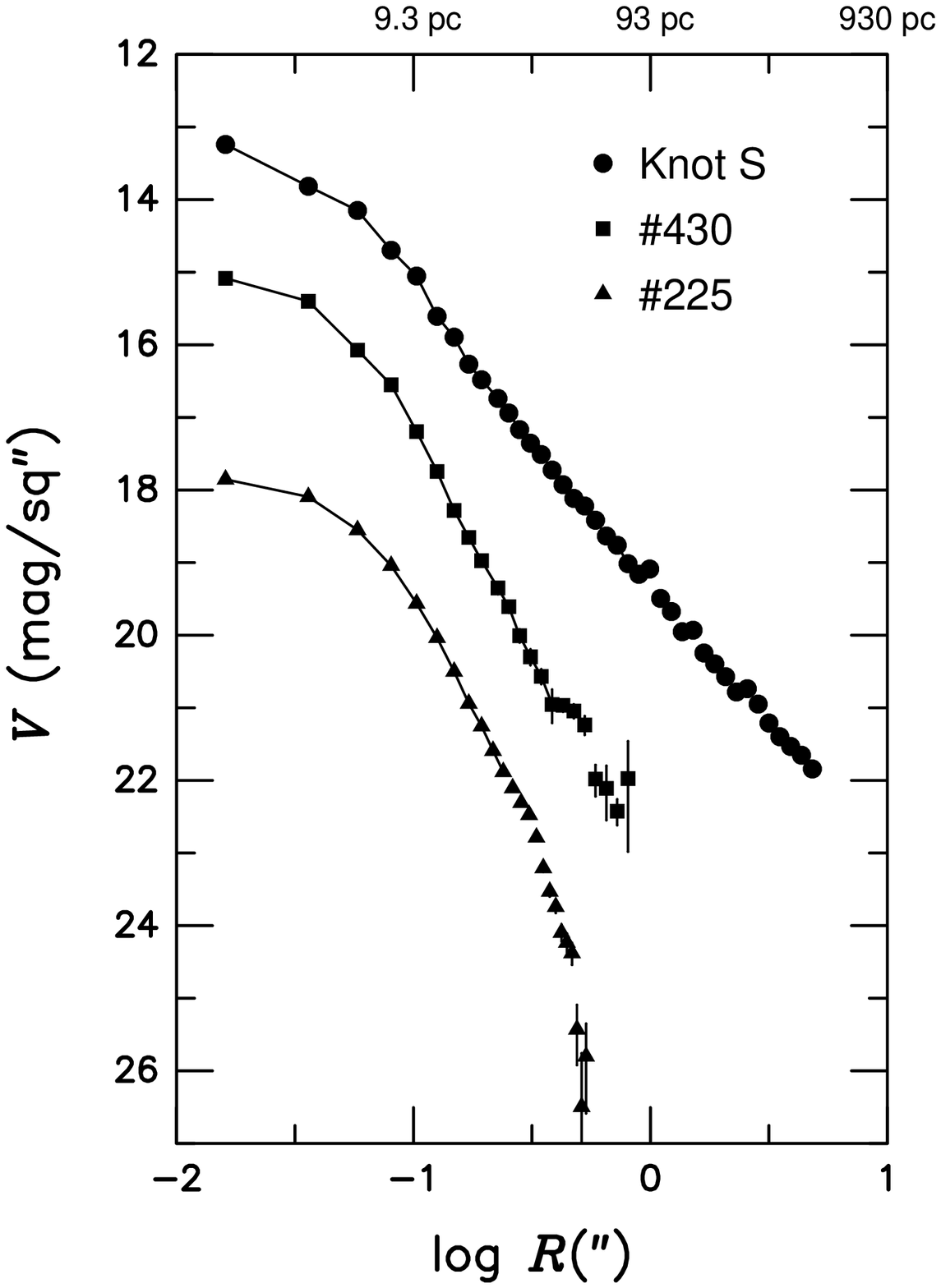,height=6.5cm,angle=0}
}}
\caption{
The Antennae galaxies (NGC 4038/39), and radial brightness profiles of
three clusters in them (Whitmore et al.\ 1999).
}
\label{fig1}
\end{figure}

As an example of such evolution, Fig.~\ref{fig1}b shows the radial-brightness
profiles of three massive clusters in NGC 4038/39:  Knot S and \#430 are
both very young (7 and 11~Myr) and display pure power-law envelopes, while
the older cluster \#225 ($\sim$500~Myr) shows both a larger core and an
envelope with a distinct tidal cutoff.  This suggests that young clusters
are born with power-law envelopes that then get truncated by external tidal
forces.

The power-law shape of the luminosity function appears to be a universal
property of young cluster systems in starburst galaxies, irrespective of
the exact cause of the burst (e.g., Meurer et al.\ 1995).  The
similarities between this power law and the power-law mass function of giant
molecular clouds---including the similar observed mass ranges---strongly 
suggest that young clusters form from giant molecular clouds suddenly
squeezed by a rapid increase in the pressure of the surrounding gas,
as encountered especially in shocks (Jog \& Solomon 1992; Harris \&
Pudritz 1994; Elmegreen \& Efremov 1997).  This is perhaps the single
most important result to have emerged from observations of young
clusters in ongoing mergers and starburst galaxies.

\section{Globular Clusters in Young Merger Remnants}

Studying GCs in young (0.3\,--\,1 Gyr) merger remnants offers two main
advantages: (1) Dust obscuration is less of a problem than in ongoing
mergers, and (2) most compact bright sources are true globular clusters. 
The second fact follows from the clusters' measured half-light radii and
ages.  These ages typically exceed 100~Myr, or $\sim$25\,--\,50 internal
cluster-crossing times $t_{\rm cr}$, and thus indicate that such clusters
are gravitationally bound.  In contrast, most clusters in ongoing mergers
like NGC 4038/39 and NGC 3256 are $\la$30~Myr or $\la$10$\,t_{\rm cr}$ old
and may eventually disperse.  Therefore, the time lapse between the peak of
cluster formation and the completion of a merger helps separate the wheat
from the chaff.

\begin{figure}[t]
\centerline{\hbox{
\psfig{figure=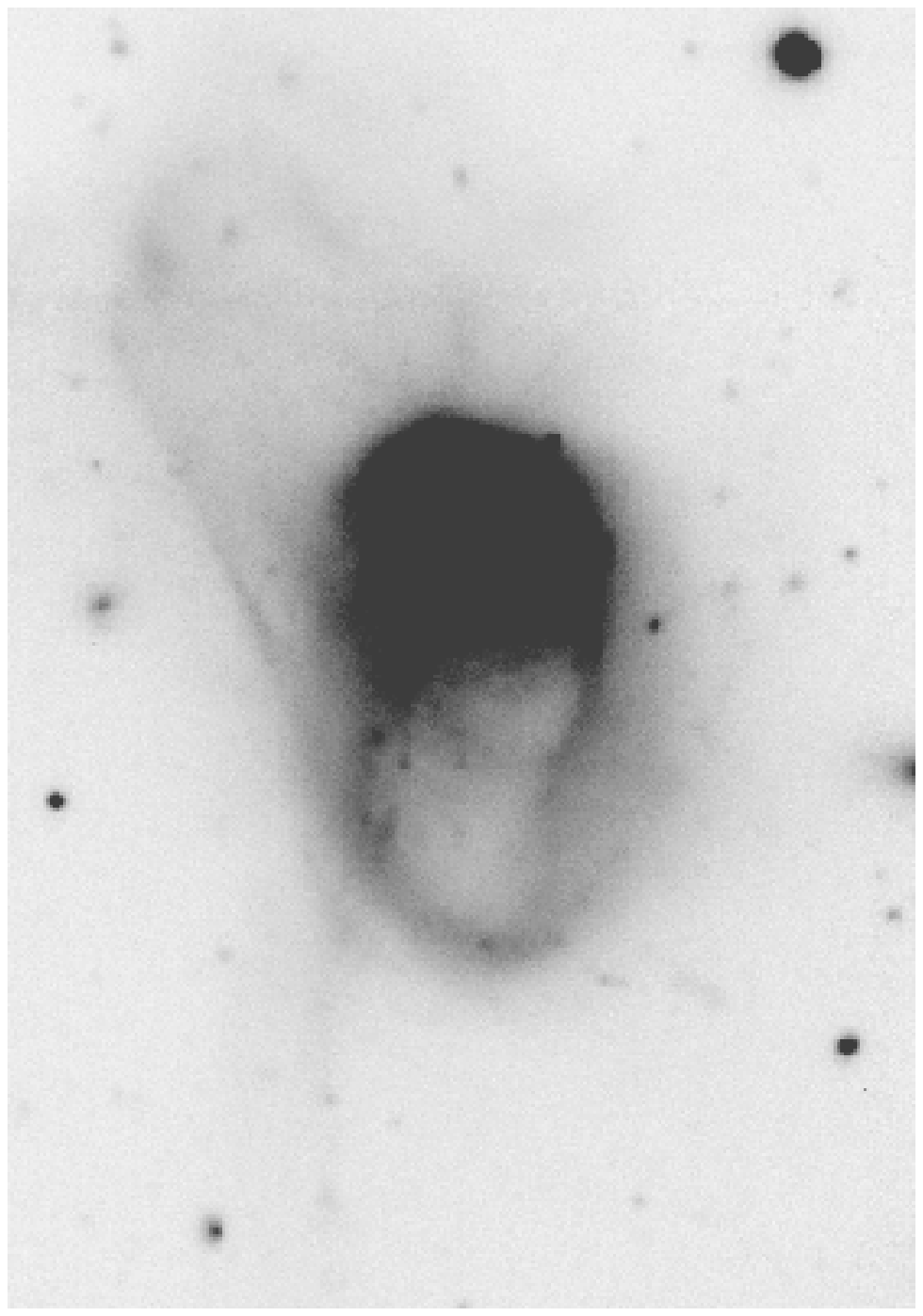,height=5cm}
%\hskip 0.1truecm
\psfig{figure=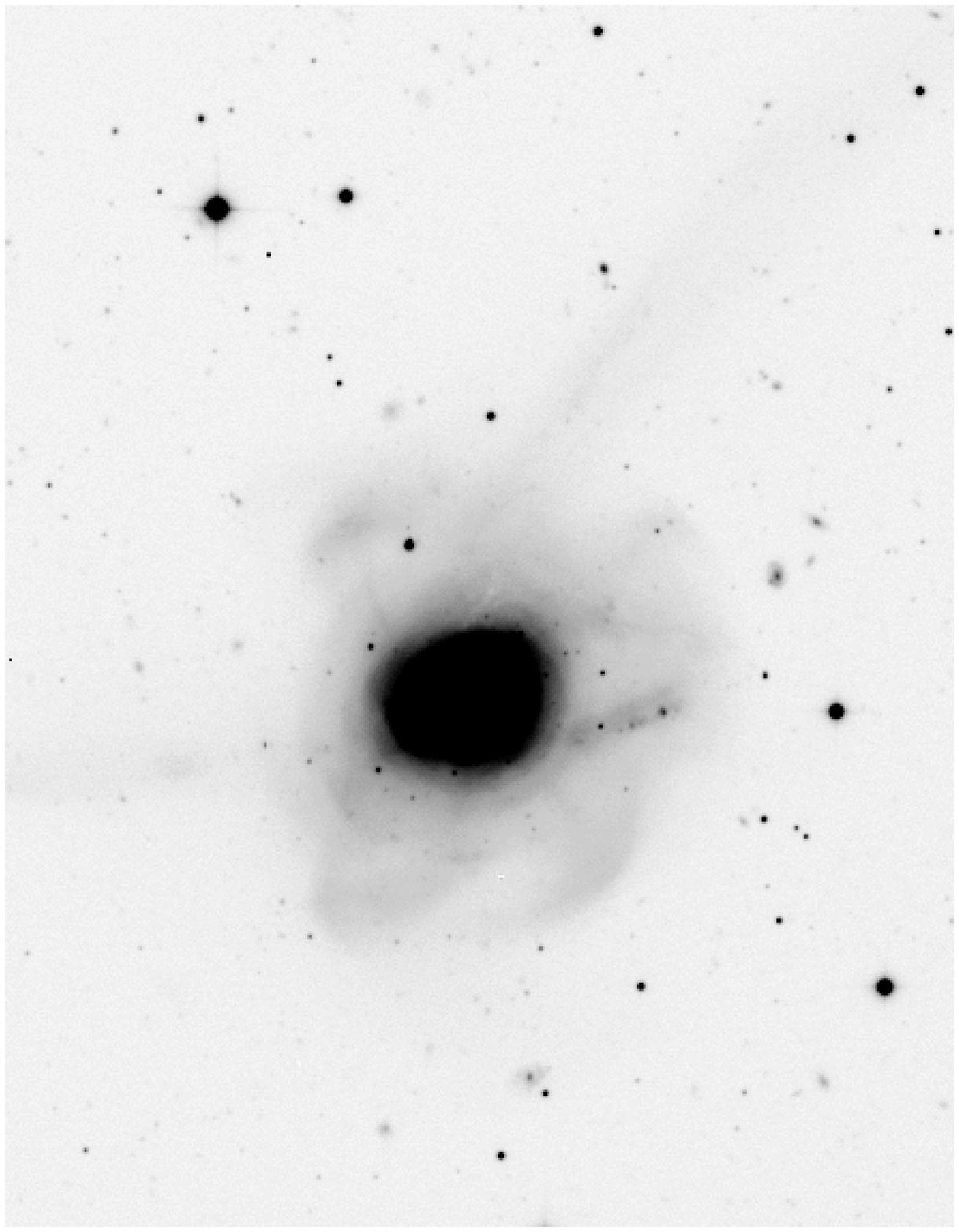,height=5cm}
%\hskip 0.1truecm
\psfig{figure=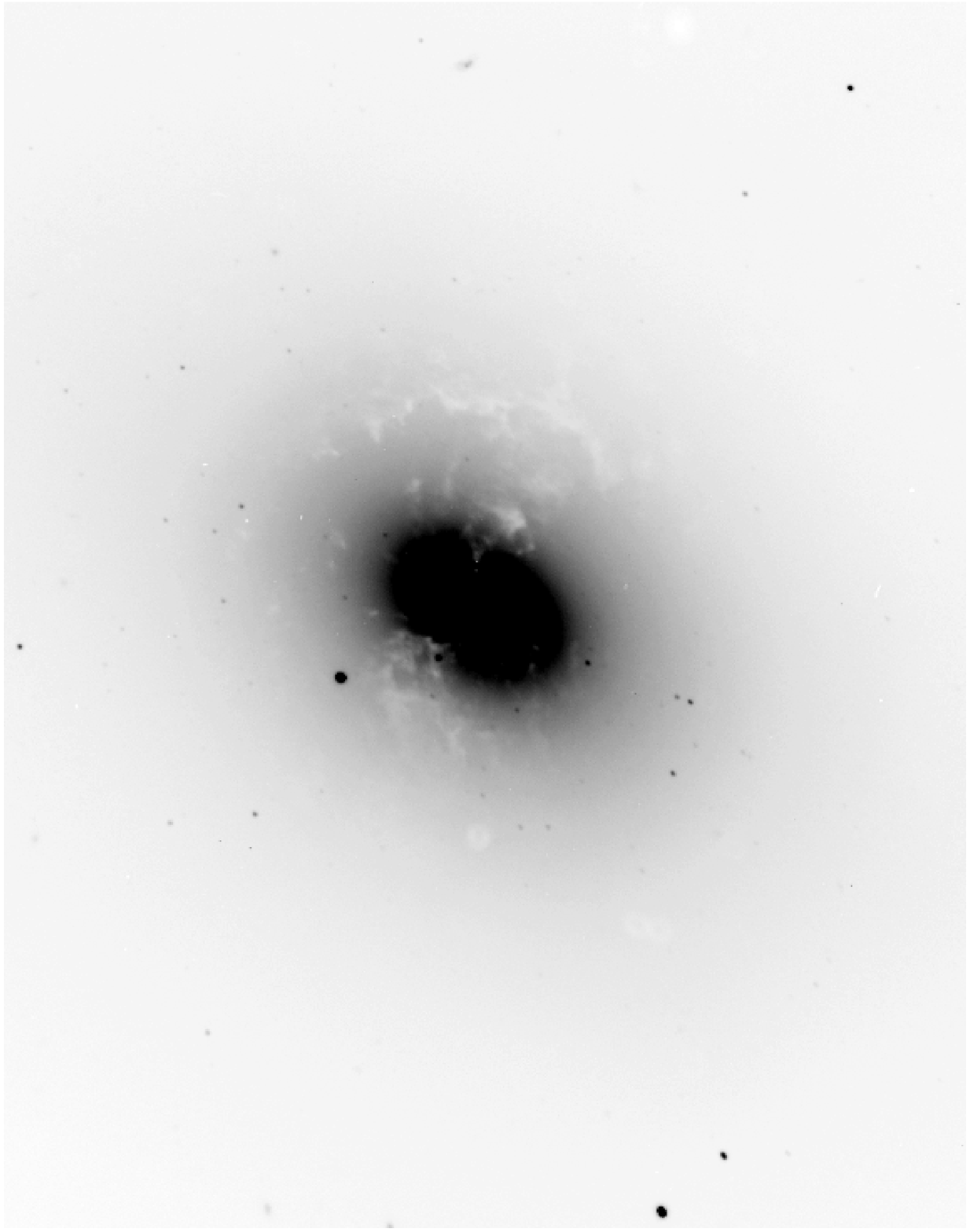,height=5cm}
}}
\caption{
Three galaxies with young and intermediate-age globulars: (a) NGC 3921 and
(b) NGC 7252 feature many young ($\sim$0.3\,--\,0.6~Gyr) halo GCs of
near-solar metallicity; (c) NGC 1316 possesses a subpopulation of
intermediate-age ($\sim$3~Gyr) GCs, also of near-solar metallicity.
}
\label{fig2}
\end{figure}

Globular-cluster systems have been studied with {\it HST}\, in several
young merger remnants, including NGC 1275 (Carlson et al.\ 1998), NGC 3597
(Carlson et al.\ 1999), NGC 3921 (Schweizer et al.\ 1996), and NGC 7252
(Miller et al.\ 1997).  Each of these remnants hosts about
\mbox{10$^2$\,--\,10$^3$} compact sources that appear to be luminous young
GCs.  Age dating based on broad-band photometry shows that the majority of
these globulars formed in a relatively short, 100\,--\,200~Myr time span
{\it during}\, each merger.  The young GCs appear strongly concentrated
toward their host galaxies' centers, half of them lying typically within
$\la$5~kpc from the nucleus.  In addition to its 102 candidate GCs,
NGC 3921 (Fig.~\ref{fig2}a) also hosts about 50 fuzzier objects that are
likely stellar associations.  These associations have colors ranging from
relatively blue to quite red and may be in the process of dispersing.
Interestingly, only three of these fuzzy objects lie within the central
5~kpc, presumably because most associations were too fragile to survive the
intense churning near the merger's center.

%%%  Parameters in this two-panel figure are very interconnected; please
%%%  do NOT easily modify any of them!
\begin{figure}[t]
\centerline{\hbox{
\psfig{figure=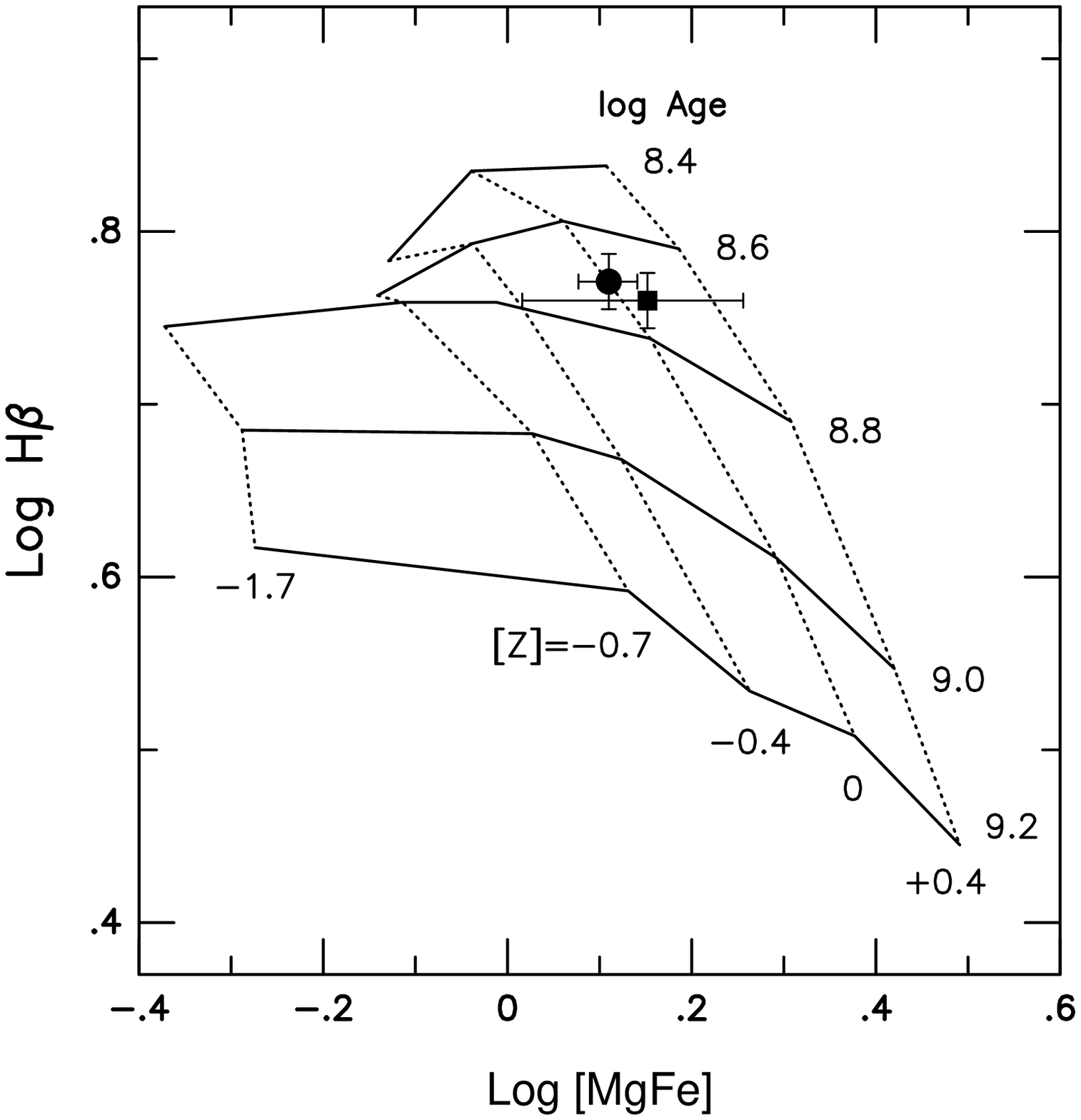,width=5.6cm}
\psfig{figure=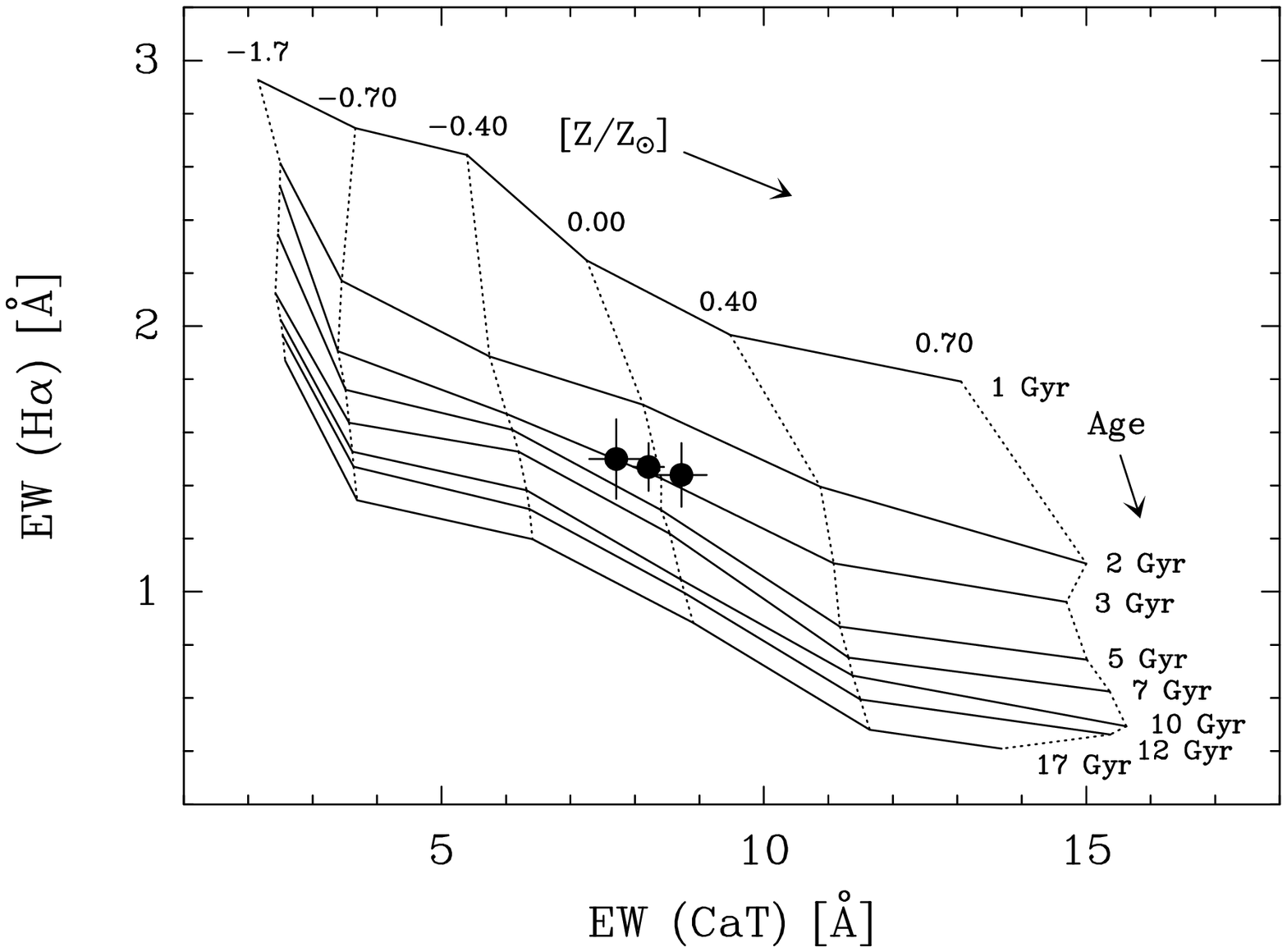,width=7.4cm,angle=270}
}}
\caption{
(a) {\it (left)} H$\beta$\,--\,[MgFe] diagram for two GCs in NGC 7252
(Schweizer \& Seitzer 1998), and
(b) {\it (right)} H$\alpha$\,--\,Ca-triplet diagram for three GCs in NGC 1316
(Goudfrooij et al.\ 2001a).  The NGC~7252 clusters are 550$\,\pm\,$50~Myr
old, the NGC~1316 clusters 3.0$\,\pm\,$0.5~Gyr.  Note that both sets of
clusters have near-solar metallicities.
}
\label{fig3}
\end{figure}

The GC system of NGC 7252 (Fig.~\ref{fig2}b) has been studied in some
detail.  Spectroscopy shows that seven of eight young GCs
feature strong Balmer absorption lines (EW[H$\beta$]\,= 6\,--\,13~\AA)
indicative of a main-sequence turnoff dominated by A-type stars.  The
GCs' age distribution appears very peaked, with six clusters having
ages in the narrow range 400\,--\,600~Myr (Schweizer \& Seitzer 1998).
Infrared photometry in the $K$-band confirms that most young globulars in
the halo of NGC 7252 are presently in the AGB phase-transition stage, which
lasts from $\sim$200~Myr to 1~Gyr (Maraston et al.\ 2001). 

The metallicity of these young halo GCs is near solar.
Figure~\ref{fig3}a shows a H$\beta$\,--\,[MgFe] diagram for the two
globulars NGC 7252:W3 and W6 ({\it data points}), from which $[Z] = 0.00\pm
0.08$ for W3 and $+0.10\pm 0.17$ for W6.  A fascinating object is S101, a
{\it freshly born halo cluster}\, located in an H$\;$II region that is
falling back into NGC 7252 from a tidal tail (with $\Delta v_{\rm rad} =
-241$~km~s$^{-1}$) and has a metallicity of $[Z] = -0.12\pm 0.05$.  This
cluster, located at a projected distance of 15~kpc, suggests
that young GCs can form with considerable time delays when tidally
ejected gas crashes back into a remnant.  

The line-of-sight velocity dispersion of the eight spectroscopically observed
GCs in NGC~7252 is $140\pm 35$~km~s$^{-1}$, leaving little doubt that
these clusters belong to a halo population.  {\it HST\,} photometry shows
that there are $\sim$300 similar {\it young} GCs in the halo, in addition
to the {\it old} GCs that likely belonged to the halos of the two input
spirals.  Hence, the color distribution of the GCs is bimodal, with the
main peak at $(V-I)_0 \approx 0.65$\, due to the clusters formed
400\,--\,600~Myr ago and the secondary peak at $(V-I)_0 \approx 0.95$\,
likely due to the brightest of the old metal-poor GCs (Miller et al.\ 1997).  

In short, the merger of two gas-rich spirals in NGC 7252 has led to a young
remnant with a bimodal population of halo globulars.  Besides the universal
old metal-poor GCs the halo also features many second-generation GCs that
are {\it young}\, and {\it metal-rich}.  The situation appears to be similar
in the young remnants NGC 3597 and NGC 3921, and perhaps also in NGC 1275.
The many properties that these remnants share with ellipticals suggest not
only that the remnants are present-day protoellipticals (e.g., Schweizer
1998), but also that E and S0 galaxies with bimodal cluster distributions
may have formed {\it their}\, second-generation metal-rich GCs in a similar
manner.  Interestingly, the ratio of young to old GCs is $\ga$0.4 in NGC 3921
(Schweizer et al.\ 1996) and $\sim$0.7 in NGC 7252 (Miller et al.\ 1997).
These ratios compare well with the {\it mean} ratio of $\sim$0.6 for
metal-rich/metal-poor GCs observed in normal giant ellipticals.

\section{Globular Clusters in Intermediate-Age Merger Remnants}

If indeed E and S0 galaxies with bimodal cluster distributions formed through
mergers similar to those described above, we should be able to find such
galaxies with second-generation GCs of intermediate age (1\,--\,7 Gyr).
These galaxies could help us trace the evolution of second-generation
GC systems from young through intermediate to old age.  Potential tracers
of such evolution are, e.g., the GC color distribution, luminosity function,
and radial distribution.

There is now evidence for the presence of intermediate-age GCs in about
ten elliptical galaxies, including NGC 1316, 5128, 1700, 3610,
4365, and 6702.  The best case is NGC 1316, where spectra of
GCs support the intermediate ages found from broad-band photometry
(Goudfrooij et al.\ 2001a,\,b).  Figure~\ref{fig3}b shows measured
equivalent widths of H$\alpha$ for three bright GCs plotted versus
the equivalent width of the Ca$\:$II triplet. From the superposed model
grid one can see that all three GCs are about $3.0\pm 0.5$~Gyr old and have
close to solar abundances.  Their ages agree with the ages inferred from $BVI$
and $JHK$ photometry for 50\,--\,60\% of a sample of $\sim$300 GCs in this
galaxy.  Therefore, the red peak of the bimodal color distribution in
NGC 1316 clearly contains GCs of intermediate age, and this merger
remnant provides a valuable evolutionary link between young remnants like
NGC 7252 and old ellipticals with bimodal cluster distributions.

\begin{figure}[t]
\centerline{\psfig{figure=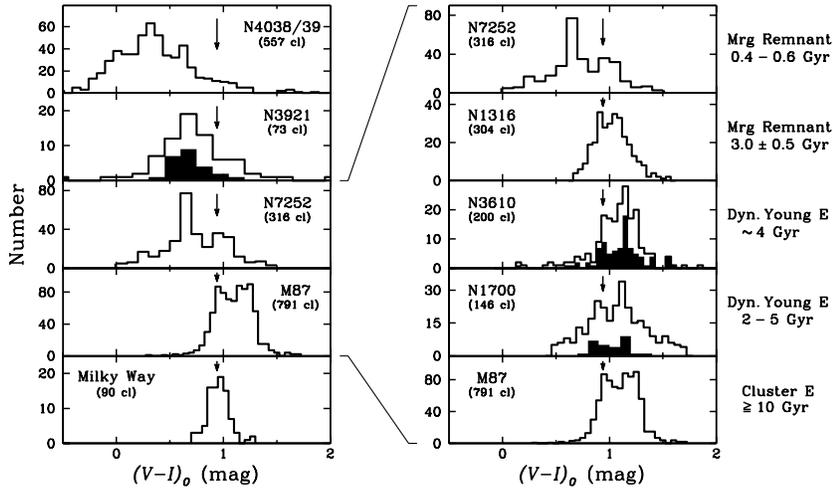,width=11.0cm,angle=-90}}
\caption{
Color distributions of GCs in 3 mergers and 4 ellipticals, compared with
Milky Way.  Location of old metal-poor GCs is indicated by arrows.  Note
color evolution of 2nd-generation GCs from very blue in NGC\,4038/39 to red
in M87.  The right panels show that the color crossover occurs between
0.5~Gyr and 3~Gyr (GC ages given in margin).
}
\label{fig4}
\end{figure}

For the other galaxies with candidate intermediate-age GCs
we have to rely on broad-band colors.  Model simulations of bimodal GC
populations with second-generation clusters of solar metallicity predict
what we can expect to observe at different ages (Whitmore et al.\ 1997,
esp.\ Fig.\ 15): At 0.5~Gyr the second-generation GCs should appear both
bluer and $\sim$2~mag brighter than the old metal-poor GCs, as is observed
in the young remnants discussed above.  Then the aging second-generation GCs
become redder.  At 1.0\,--\,1.5~Gyr they reach about the same $V-I$ color as
old GCs, but are still $\sim$1.5~mag brighter.  At 3~Gyr they are already
distinctly redder than the old GCs but still 0.5\,--\,1~mag brighter, while
at $\ga$10~Gyr they appear both redder and slightly fainter.
Figure~\ref{fig4} illustrates that this predicted crossover of GC colors
does indeed occur.  Shown are the color distributions of clusters in seven
galaxies, with second-generation GCs ranging from very young and blue in The
Antennae to old and red in M87.  Whereas the general evolution from blue to
red colors for second-generation GCs has been known for some time, the
more detailed transition shown in the right-hand panels of Fig.~\ref{fig4} is
new.  The new data for NGC 1316 (Goudfrooij et al.\ 2001b), NGC 3610
(Whitmore et al.\ 2002), and NGC 1700 (Brown et al.\ 2000) diminish the gap
in known cluster ages from 0.5\,--\,10$^+$~Gyr previously to a current
$\sim$0.5\,--\,3~Gyr.

A promising new method for finding more E\,+\,S0 galaxies with intermediate-age
globulars is to break, or at least diminish, the age--metallicity degeneracy
by supplementing {\it HST\,} photometry in $VI$ with ground-based photometry
in the $K$ band.  Applying this method to NGC 4365 (E3), Puzia et
al.\ (2002) find a significant population of very metal-rich GCs of
intermediate age in addition to old metal-poor and old metal-rich populations.

\begin{figure}[t]
\centerline{\psfig{figure=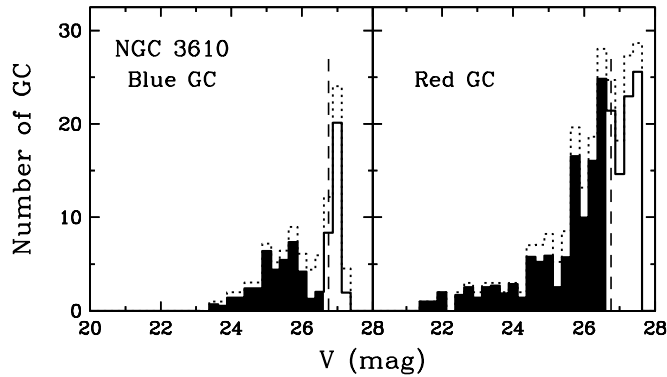,height=5.0cm}}
\caption{
Luminosity functions of blue and red GCs in NGC 3610. Half-completeness
limits ({\it vertical dashed}) mark boundary up to which the LFs can be
trusted.  {\it Shaded\,} areas mark background-corrected LFs, {\it dotted
lines\,} uncorrected LFs.  Notice the nearly lognormal shape of LF for
blue GCs and power-law shape for red GCs (Whitmore et al.\ 2002).
}
\label{fig5}
\end{figure}

The luminosity function (LF) of second-generation GCs is another potential
tracer of systemic evolution.  The transition from the power-law form
observed in young cluster systems to the log-normal form observed in old
GC systems has been predicted theoretically by Fall \& Zhang (2001).
It is a consequence of the preferential disruption of low-mass clusters by
various mechanisms, of which the main one is internal two-body relaxation
and evaporation.  The resulting erosion of the low-mass end should be evident
in the observed LFs of second-generation GC systems that form an age sequence. 

Striking differences between the LFs of red, metal-rich GCs and blue,
metal-poor GCs have indeed been found in two bona fide ellipticals: NGC 1316
and NGC 3610. Both galaxies show fine structure indicative of mergers
involving disks during the past few Gyr.  Figure~\ref{fig5} compares the
observed LFs of the blue ($0.8\leq V-I\leq 1.02$) and red($1.02 < V-I
\leq 1.3$) GCs in NGC 3610.  Whereas the LF of the blue clusters is nearly
lognormal, as expected for old GCs, that of the red cluster is well
approximated by a power law of index $-1.78\pm 0.05$ (Whitmore et al.\ 2002).
Thus, it supports the notion that many of the red GCs are not ancient, but
formed relatively recently.  The turnover predicted by Fall \& Zhang's models
for 3\,--\,4~Gyr old clusters is presently not detected, perhaps because it
lies near the 50\% completeness limit.  Therefore, an effort is under way to
obtain new, still deeper observations with {\it HST\,} and the ACS camera.

The red GCs of NGC 1316 appear to have a power-law LF as well, though
Goudfrooij et al.\ (2001b) at first found an exponent of $-1.23\pm 0.26$.
A reanalysis by these authors shows, however, that a calculational error
was made and the true value is $-1.7\pm 0.1$.  Hence, in both NGC 1316 
and NGC 3610 power-law LFs support the notion that the red GCs are of
intermediate age.

\begin{figure}[t]
\centerline{\psfig{figure=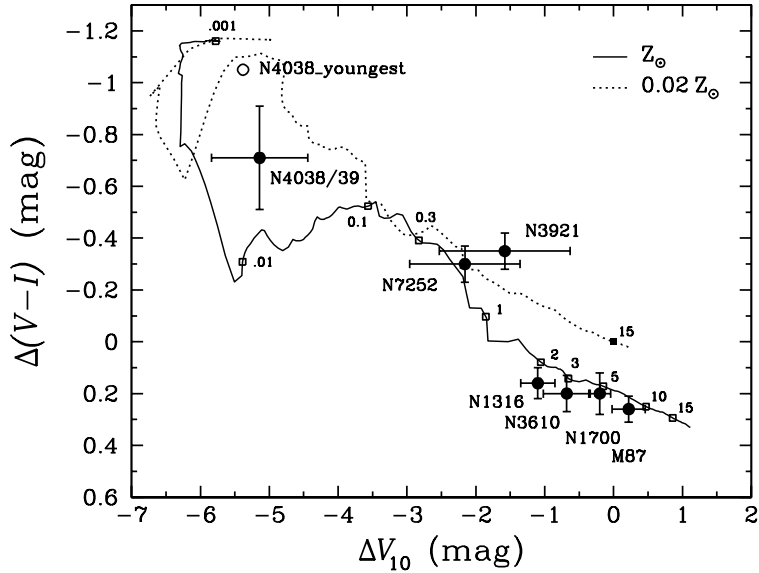,width=10.0cm,angle=270}}
\caption{
$\!\Delta(V\!-\!I)$\,--\,$\Delta V_{10}$ diagram showing color difference
between 2nd-generation and old GCs plotted vs.\ magnitude difference
between 10th-brightest 2nd-generation GC and its old counterpart. Points
with error bars show values for GC systems of 3 merger galaxies and 4 Es.
Lines give evolutionary tracks for model clusters of solar and 1/50th solar
metallicity (Bruzual \& Charlot 1996), and are marked with cluster ages
in Gyr.  Note that the observed seven GC systems form an age sequence.
(After Whitmore et al.\ 1997, with some new data added.)
}
\label{fig6}
\end{figure}

A powerful tool for displaying evolutionary trends of GC systems as a
function of age is the $\Delta(V\!-\!I)$ vs.\ $\Delta V_{10}$ diagram,
which combines color and luminosity information (Whitmore et al.\ 1997).
Figure~\ref{fig6} shows a version of this diagram in which $\Delta(V\!-\!I)$,
the reddening-corrected color difference between the peaks due to
second-generation GCs and to old metal-poor GCs, is plotted versus
$\Delta V_{10}\,$, the magnitude difference between the 10th-brightest
second-generation GC and the 10th-brightest old GC.  Data points with error
bars mark the locations of the GC systems for the galaxies of Fig.~\ref{fig4}. 
Note that the seven GC systems lie roughly along the evolutionary track for
second-generation model clusters of solar metallicity ({\it solid line}).
This supports the notion that most second-generation GCs are relatively
metal-rich ($[Z]\approx -0.8$ to $+$0.2), as verified spectroscopically
for GCs in NGC 7252 (Schweizer \& Seitzer 1998), NGC 1316 (Goudfrooij et
al.\ 2001a), and M87 (Cohen, Blakeslee, \& Ryzhov 1998).  But above all,
the $\Delta(V\!-\!I)$\,--\,$\Delta V_{10}$ diagram demonstrates quite clearly
that the GC systems of the three merger galaxies and four ellipticals form
an {\it age sequence}.

Along this sequence, NGC 1316, 3610, and 1700 seem to be good candidates for
the kind of descendants of merged galaxy pairs that Ivan King was asking
Toomre about at Yale (see \S 1).  With new search techniques and
astronomers' growing interest in the subject, we can---during the next
5\,--\,10 years---expect the discovery of many more transition objects
in what I call the ``King Gap.'' Hopefully, some of
these galaxies and their GC systems will fill in the presently still
empty age range of about 0.5\,--\,3~Gyr.  Realistically, however, we
must face the fact that it will be difficult to find objects in the
age range 1\,--\,2~Gyr, where intermediate-age metal-rich clusters blend
in $V-I$ color with the old metal-poor clusters.  In this age range,
time-consuming spectroscopy will be required to separate the cluster
subpopulations of different age and metallicity.

Nevertheless, the GC subsystems of intermediate age now known in about
ten ellipticals do seem to form an evolutionary link between the young
metal-rich GCs observed in recent merger remnants and the old metal-enriched
GCs found in most giant ellipticals.  This, then, seems to be a fitting
present from us all to Ivan King on his 75th birthday!

\section{Globular-Cluster Formation and Old Ellipticals}

Given the observed propensity of globular clusters to form in the high-pressure
environments of merger-induced starbursts (Schweizer 1987; Ashman \& Zepf
1992; Jog \& Solomon 1992; Elmegreen \& Efremov 1997), GCs in old ellipticals
($>$7~Gyr) serve as valuable fossils of these galaxies' early star-formation
history.  The discovery of {\it bimodal GC populations} in nearby
ellipticals (Zepf \& Ashman 1993; Whitmore et al.\ 1995) has convinced many
sceptics that major mergers played a role in forming at least the metal-rich,
second-generation GCs, though alternative formation scenarios have been
proposed as well.  In a landmark study of M49 (= NGC 4472), Geisler et al.\
(1996) discovered that the metal-rich clusters lie on average closer to the
center than the metal-poor ones, as predicted by Ashman \& Zepf's merger
model.  This solved the old puzzle of why some GC systems show steeper radial
abundance gradients than their host ellipticals, a fact attributable in
M49 to the radially varying ratio of metal-rich to metal-poor GCs.  Mergers
involve much gaseous dissipation, which explains quite naturally the stronger
central concentration of second-generation clusters.

{\it Kinematic differences} between metal-poor and metal-rich globular
clusters in E and S0 galaxies also seem to point to major mergers in the
past history of these hosts.  Both the outwardly increasing mean rotation
of GCs in M87 (Kissler-Patig \& Gebhardt 1998) and flips in the mean-rotation
axis of the clusters as a function of radius (C\^ot\'e et al.\ 2001)
are difficult to explain in any monolithic-collapse model, but are a natural
consequence of major mergers.   In such mergers, old, pre-existing clusters
from the halos of the input galaxies can be expected to acquire large
velocity dispersions and significant outer rotation stemming from the
galaxies' orbital angular momentum.  Second-generation clusters formed during
the merger(s), on the other hand, should show lower velocity dispersions
and less net rotation, as observed in M49 (Zepf et al.\ 2000).

Even in the {\it radial distributions}\, of GCs within their host galaxies we
may begin to see an evolutionary sequence from young to old merger remnants.
In young remnants (e.g., NGC 3921, 7252) the radial distribution of
second-generation GCs is virtually identical to that of the galaxy
light.  This indicates that the young GCs and their progenitors experienced
the same violent relaxation as did the average star, suggesting that the GC
progenitors were relatively compact {\it giant molecular clouds}\, orbiting
among the disk stars of the input spirals (Schweizer et al.\ 1996).  There
is tentative evidence for subsequent central erosion of GC
systems, presumably due to tidal shocking of GCs during passages close to the
center:  At $r = 1.2$~kpc in NGC 1316, the radial distribution of GCs shows a
deficit of $\sim$45\% relative to the integrated star light (Goudfrooij et
al.\ 2001b), while in old cluster ellipticals the corresponding deficit is
significantly larger (e.g., Capuzzo-Dolcetta \& Donnarumma 2001).  Therefore,
there appears to be a continuum of radial distributions of GCs ranging from
young-cluster distributions as strongly centrally concentrated as the host
merger remnants to old-cluster distributions typically less concentrated
than the host ellipticals. It remains to be seen whether this tentative
dynamical sequence will be confirmed as further examples of intermediate-age
GC systems are added to the sample.

Perhaps the single most challenging question concerning GCs is why the
old metal-poor GCs are so universally similar in all types of galaxies
and environments. I believe that observations of cluster formation in
present-day mergers and starburst galaxies have yielded a crucial clue:
Whenever an ensemble of giant molecular clouds is exposed to a rapid pressure
increase, a significant fraction of these clouds get shocked and turn into GCs
(Jog \& Solomon 1992).  The question then is:  Was there a {\it universal}\,
pressure increase early in the history of the universe that might explain the
surprisingly uniform ages and properties of old metal-poor GCs?

The cosmological reionization at $z$~$\approx$ 7\,--\,15 may have provided
just such a universal pressure increase, which in turn led to the
near-synchronous formation of metal-poor GCs from early giant molecular
clouds (Cen 2001).  If this hypothesis is correct, the following
{\it unified scenario of GC formation}\, emerges.

Most globular clusters in the universe formed from shocked giant molecular
clouds.  The first-generation GCs formed near-simultaneously from pristine
such clouds shocked by the strong pressure increase accompanying
cosmological reionization.  They populate all types of galaxies from
dwarfs through spirals and ellipticals to giant cDs. Later-generation
(``second-generation'') GCs formed during subsequent mergers from
metal-enriched giant molecular clouds present in the merging disks.
Major mergers, some of which occur to the present time, led to elliptical
remnants with a mixture of first- and second-generation GCs revealed by
their bimodal color distributions.  Minor mergers tended to form S0 galaxies
and early-type spirals, again with a mixture of first- and second-generation
GCs.  However, unlike in ellipticals many of the second-generation metal-rich
GCs in S0 galaxies may belong to a thick-disk population if they stem mainly
from giant molecular clouds that belonged to the dominant input disk.
Finally, a minority of ``second-generation'' GCs form sporadically from
occasional pressure increases in calmer environments, such as in interacting
irregulars or barred spirals.

\acknowledgements
%I thank Paul Goudfrooij, Pat Seitzer, and Brad Whitmore for their kind
%permission to reproduce figures. 
%I also gratefully acknowledge support from NASA through various {\it HST}\,
%grants and from the NSF through Grants AST-99\,00742 and AST-02\,05994.
I thank Paul Goudfrooij, Pat Seitzer, and Brad Whitmore for permission to
reproduce figures, and acknowledge support from the NSF through Grants
AST-99\,00742 and AST-02\,05994 and from NASA.

%%%%%%%%%%%%%%%%%%%%%%%%%%%%%   REFERENCES   %%%%%%%%%%%%%%%%%%%%%%%%%%%%%%%%%

\end{document}